\documentclass[journal]{IEEEtran}

\usepackage{amsmath,epsfig,dsfont}
\usepackage{subfigure}
\usepackage{algorithm}
\usepackage{algorithmic}
\usepackage{amssymb,bm}
\usepackage{amsfonts}
\usepackage{stfloats}
\usepackage{multirow}
\ifCLASSINFOpdf
\else
\fi
\begin{document}

\title{Evolutionary Information Diffusion over Social Networks}
%
%
%
\author{\authorblockN{Chunxiao~Jiang\authorrefmark{1}\authorrefmark{2}, Yan~Chen\authorrefmark{1}, and K. J. Ray~Liu\authorrefmark{1}} \\ 
      \small\authorblockA{\authorrefmark{1}Department of Electrical and Computer Engineering, University of Maryland, College Park, MD 20742, USA\\ 
          \authorrefmark{2}Department of Electronic Engineering, Tsinghua University, Beijing, 100084, P. R. China\\ 
        E-mail:\{jcx, yan, kjrliu\}@umd.edu}}
\maketitle

\begin{abstract}
Social networks have become ubiquitous in our daily life, as such it has attracted great research interests recently. A key challenge is that it is of extremely large-scale with tremendous information flow, creating the phenomenon of ``Big Data''. Under such a circumstance, understanding information diffusion over social networks has become an important research issue. Most of the existing works on information diffusion analysis are based on either network structure modeling or empirical approach with dataset mining. However, the information diffusion is also heavily influenced by network users' decisions, actions and their socio-economic connections, which is generally ignored in existing works. In this paper, we propose an evolutionary game theoretic framework to model the dynamic information diffusion process in social networks. Specifically, we analyze the framework in uniform degree and non-uniform degree networks and derive the closed-form expressions of the evolutionary stable network states. Moreover, the information diffusion over two special networks, Erd\H{o}s-R\'enyi random network and the Barab\'asi-Albert scale-free network, are also highlighted. To verify our theoretical analysis, we conduct experiments by using both synthetic networks and real-world Facebook network, as well as real-world information spreading dataset of Twitter and Memetracker. Experiments shows that the proposed game theoretic framework is effective and practical in modeling the social network users' information forwarding behaviors.
\end{abstract}
%
\section{Introduction}

During the past decade, with the rapid development of the Internet and mobile technologies, social networks have become ubiquitous. Due to its diverse implication, researchers of different disciplines have been working on social networking from various perspectives \cite{Liu}. A social network is a network made of connections and interactions among social entities, which can be individuals, websites, organizations and even intelligent equipments. Typical social network examples include Facebook/Twitter networks, hyperlink networks of websites, scientific collaboration/citation networks, and Internet of Things (IoT) \cite{sns}.

\begin{figure*}[!t]
  \centering
  \centerline{\epsfig{figure=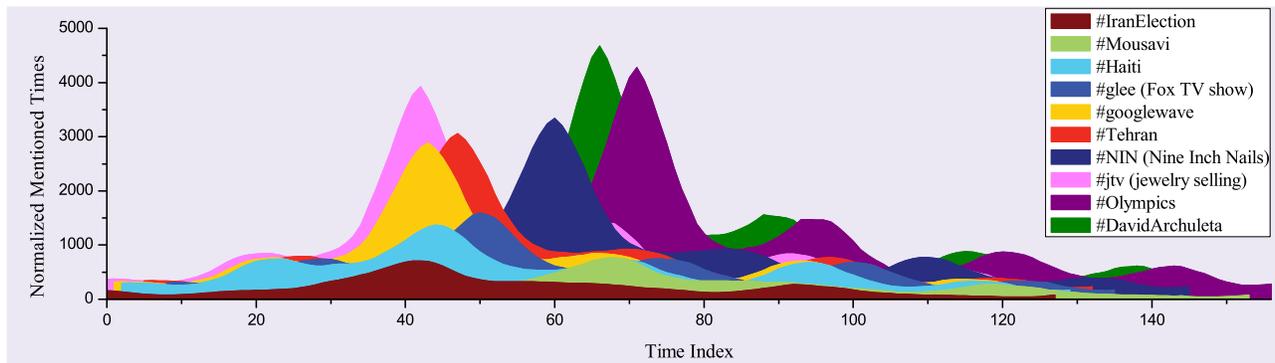,width=17cm}}
  \caption{Normalized number of forwarding of different Twitter hashtags using dataset \cite{dataset}, collected from Sep. 2008 to Aug. 2009.}\label{twitter}
\end{figure*}

Today's social networks are of very large-scale and becoming tremendous-scale, e.g., Facebook has nearly one billion active users as of September 2012 \cite{facebookuser}. Over such large-scale social networks, users are exchanging their information, status and ideas in the form of ``word-of-mouth'' communications everyday. Under such a circumstance, how the information diffuses over the underlying social network has become an important research problem. On one hand, the study of information diffusion can help the enterprises/polititians to identify the influential users and links, on which the initial advertisement/advocation will be most effective. On the other hand, from the security point of view, the study of information diffusion can also help to prevent the detrimental information spreading, e.g. computer virus. The recent developments of information and communication technologies enable us to collect, store, and access the real-world big data, making the information diffusion research versatile and meaningful, at the same time more challenging. For instance, Fig.\,\ref{twitter} shows the normalized number of forwarding of different Twitter hashtags using dataset \cite{dataset}, which contains 1000 highest total volume hashtags among 6 million hashtags from Jun to Dec 2009 \cite{twitterpaper}. From Fig.\,\ref{twitter}, we can see that different phrases reached different levels of popularity at different time. How to model such information diffusion process and predict the final stable state of the information diffusion? In this paper, we will attempt to answer these questions using graphical evolutionary game theory.

The study of information diffusion originates from the research of computer virus/epidemic spreading over network \cite{1}. One of the earliest and prominent works about information diffusion is \cite{2}, which studied the dynamics of information propagation through blogspace from both macroscopic and microscopic points of views. Subsequently, there are numerous works on the information diffusion, and here we summarize the works that are most representative and relevant to our study. The existing works about information diffusion can be classified into two categories. The first category focuses on analyzing the characteristics of information diffusion \cite{3}\nocite{4,5,6}-\cite{7}. In \cite{3}, the authors discussed how to extract the most influential nodes on a large-scale social network. Later, a community-based greedy algorithm was proposed for mining top-k influential nodes in mobile social networks \cite{4}. How to restrain the private or contaminated information diffusion was studied in \cite{5} and \cite{6} through identifying the important information links and hubs, respectively. On the other hand, how to maximize information diffusion through a network was discussed in \cite{7} by designing effective neighbours selection strategies.

The second category focuses on analyzing the dynamic diffusion process over different kinds of networks using different mathematical models \cite{8}\nocite{9,10,11,12}-\cite{13}. In \cite{8}, the authors studied how social networks affect the spread of behavior and investigated the effects of network structure on users' behavior diffusion. Rather than focusing on the behavior diffusion, Bakshy \emph{et al.} studied the role of social networks in general information diffusion through an experimental approach \cite{9}. Recently, as social networks, e.g., Facebook and Twitter, become more and more popular, some empirical analysis were conducted using large-scale datasets, including predicting the speed and range of information diffusion on Twitter \cite{10}, modeling the global influence of a node on the rate of diffusion on Twitter \cite{11}, illustrating the statistical mechanics of rumor spreading on Facebook \cite{12}. Moreover, the information diffusion on overlaying social-physical networks was analyzed in \cite{13}. The information diffusion analysis presented in this paper falls into the second category, i.e., our focus is on the analysis of dynamic diffusion process.

Most of existing works on diffusion analysis mainly rely on the social network structure and/or the collected data, while totally ignore the actions and decision making of users. However, the influence of users' decisions, actions and socio-economic connections on information forwarding also plays an important role in the diffusion process. What does that actually mean? In essence, for information to diffuse, it relies on other users to forward the information they receive. However, it is not unconditional. One has to make a decision on whether or not to do so based on many factors, such as if the information is exciting or if his/her friends are interested on it, etc. This kind of interaction can often be modeled by using game theory \cite{Yan}. Therefore, in this paper, we propose a game theoretic framework to analyze the information diffusion over social network. Specifically, we find that in essence the information diffusion process on social networks follows similarly the evolution process in natural ecological systems \cite{evol}. It is a process that evolves from one state at a particular instance to another when information is being forwarded and diffused around. Thus, we consider the graphical evolutionary game to model and analyze the social network users' information forwarding strategies and the dynamic diffusion process. 

The contributions of this paper can be summarized as follows.
\begin{enumerate}
\item We propose a graphical evolutionary game theoretic framework to model and understand information diffusion over social networks. The framework reveals the dynamics of information diffusion among users through analyzing their learning, interactions and decision making. Moreover, the framework not only can illustrate the dynamic process of the information diffusion, but also predict the final diffusion state.

\item Based on the graphical evolutionary game theoretic formulation, we analyze the dynamic diffusion process over uniform degree and non-uniform degree networks. The closed-form expressions for the evolutionary stable network states are obtained. Moreover, the information diffusion over two special networks, Erd\H{o}s-R\'enyi random network and the Barab\'asi-Albert scale-free network, are also studied.

\item We conduct experiments to validate our information diffusion analysis using both synthetic networks including uniform-degree network, Erd\H{o}s-R\'enyi random network and the Barab\'asi-Albert scale-free network, and real-world Facebook network. Moreover, we also use the real-world information spreading dataset from Memetracker and Twitter \cite{meme} to further verify the diffusion analysis.
\end{enumerate}

\begin{figure*}[!t]
\begin{minipage}[t]{1\linewidth}
  \centering
  \centerline{\epsfig{figure=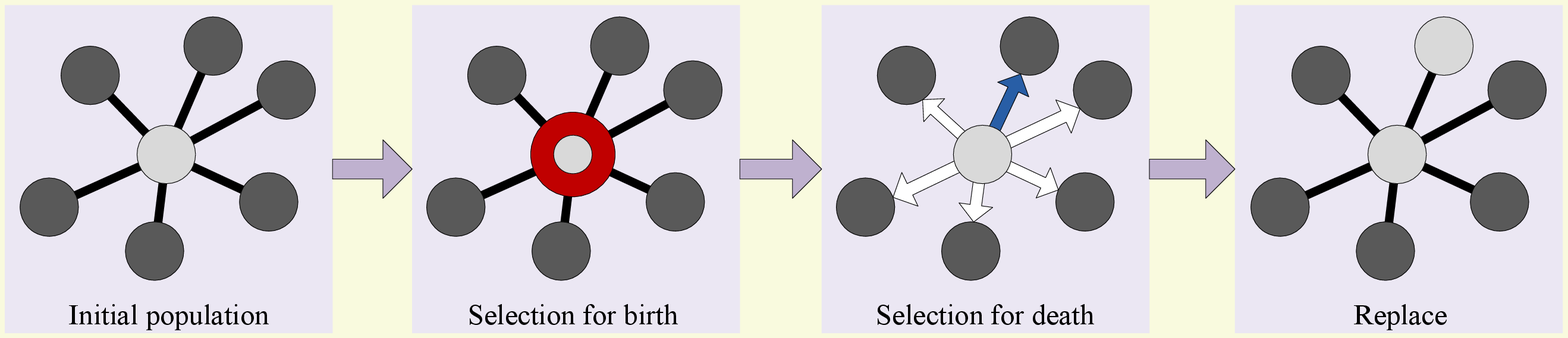,width=14cm}}
  \centerline{\scriptsize{(a) BD update rule.}}\vspace{0.3cm}
\end{minipage}
\begin{minipage}[t]{1\linewidth}
  \centering
  \centerline{\epsfig{figure=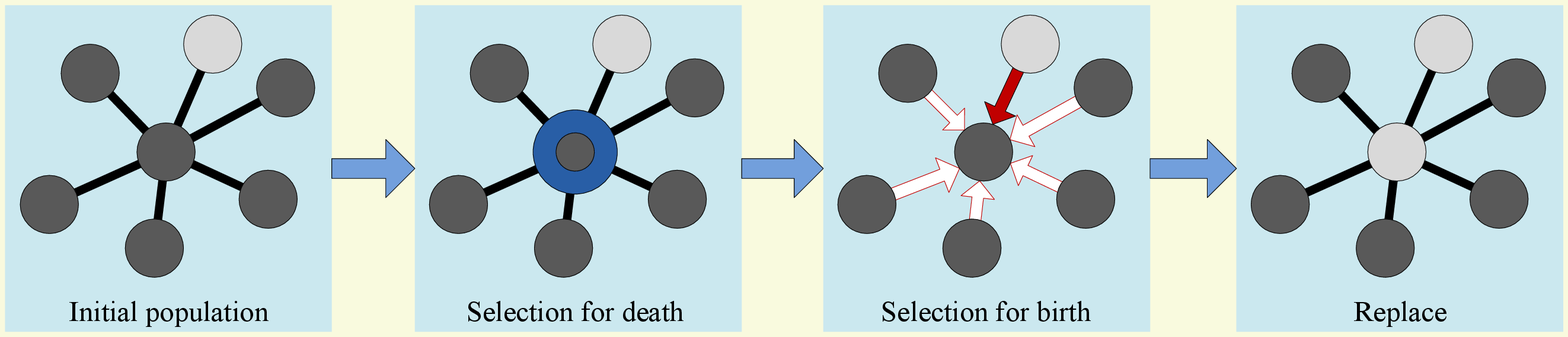,width=14cm}}
  \centerline{\scriptsize{(b) DB update rule.}}\vspace{0.3cm}
\end{minipage}
\begin{minipage}[t]{1\linewidth}
  \centering
  \centerline{\epsfig{figure=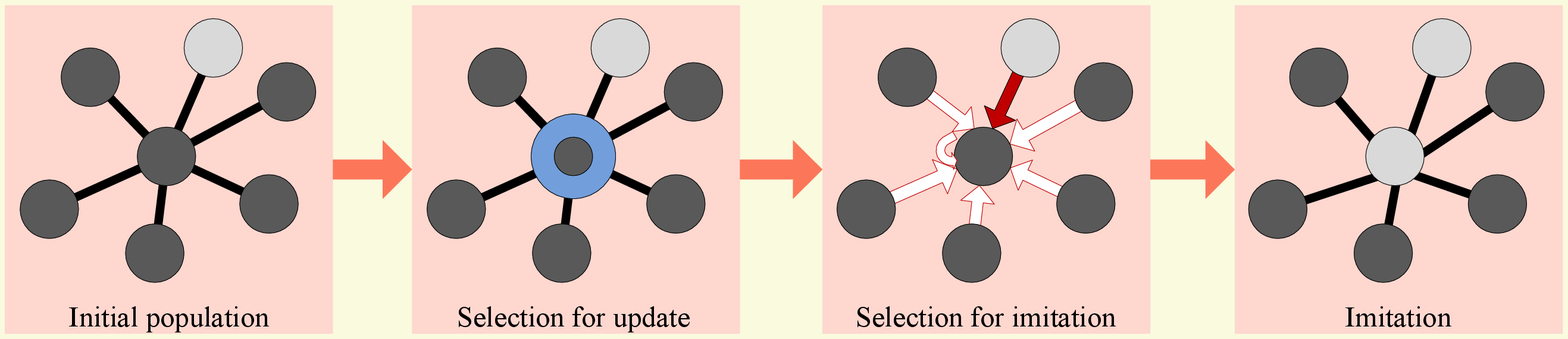,width=14cm}}
  \centerline{\scriptsize{(c) IM update rule.}}\vspace{0.2cm}
\end{minipage}
\caption{Three different update rules, where death selections are shown in dark blue and birth selections are shown in red.}\label{fig3}
\end{figure*}
The rest of this paper is organized as follows. We first give a brief overview of graphical evolutionary game formulation of information diffusion in Section II. Then, we analyze the dynamic information diffusion process over uniform and non-uniform degree networks in Section III and IV, respectively. Experiments results are shown in Section V and conclusions are drawn in Section VI.

\section{Graphical Evolutionary Game Formulation}

In this section, we discuss the formulation of the information diffusion problem using graphical evolutionary game. We first introduce the basic concepts of graphical evolutionary game, and then elaborate how to formulate the information diffusion process using graphical evolutionary game theory.

\subsection{Basic Concepts of Graphical Evolutionary Game}

The traditional interpretation of game theory is that a game with some specific rule is played among a group of static players and a static Nash equilibrium (NE) can be achieved by analyzing the players' payoff, utility function and the game rule. Evolutionary game theory (EGT) originated from ecological biology \cite{evol}, instead, imagines that a game is played over and over again by biologically or socially conditioned players who are randomly drawn from large populations \cite{evol2}. It studies the population shift and evolving process due to the influence of mutants. Meanwhile, EGT emphasizes more on the dynamics and stability of the whole population's strategies, instead of only focusing on the property of the static NE. Recently, EGT has been widely used to model users' behaviors in communication and networking area \cite{bookliu}\cite{reviewwang}, including congestion control \cite{cong}, cooperative sensing \cite{cs}, cooperative peer-to-peer (P2P) streaming \cite{p2p} and dynamic spectrum access \cite{joint}, and also image processing area \cite{yanev}. In these literatures, evolutionary game has been shown to be an effective approach to model the dynamic social interactions among users in a network.

EGT studies how a group of players converges to a stable equilibrium after a period of strategic interactions. Such a final equilibrium state is defined as the Evolutionarily Stable State (ESS). Let us consider an evolutionary game with $m$ strategies $\mathcal X=\{1,2,...,m\}$ and payoff matrix, $\mathbf U$, which is an $m\times m$ matrix, whose entry, $u_{ij}$, denotes the payoff for strategy $i$ versus strategy $j$. The system state of this game can be defined as $\mathbf p=[p_1,...,p_i,...,p_m]^T$, where $p_i$ is the fraction of population using strategy $i$ and $\sum_{i=1}^m p_i=1$. In such a case, the average mean payoff within a (sub-)population in state $\mathbf q=[q_1,q_2,...,q_m]^\prime$ against a population in state $\mathbf p$ is $\mathbf q^\prime \mathbf U \mathbf p$. A state $\mathbf p^*$ is an ESS, if and only if for all different states $\mathbf q\neq \mathbf p$, $\mathbf p^*$ satisfies follows \cite{BOMZE}:
\begin{eqnarray}
\!\!\!\!\!\!\!\!\!\!\!\!\!\!\!\!\!\!&&1)\ \mathbf q^\prime \mathbf U \mathbf p^*\le \mathbf p^{*\prime} \mathbf U \mathbf p^*,\quad\quad\quad\quad\quad\quad\quad\\
\!\!\!\!\!\!\!\!\!\!\!\!\!\!\!\!\!\!&&2)\ \mbox{if } \mathbf q^\prime \mathbf U \mathbf p^* = \mathbf p^{*\prime} \mathbf U \mathbf p^*,\quad\mathbf p^{*\prime} \mathbf U \mathbf q> \mathbf q^\prime \mathbf U \mathbf q.\label{ess}
\end{eqnarray}
The first condition guarantees that the average mean payoff of the population in a different state $\mathbf q$ against the ESS $\mathbf p^*$ does not exceed the average payoff of the population in $\mathbf p^*$, which is equivalent to the Nash equilibrium condition. The second condition shows that, in case of equality in the equilibrium condition, the average payoff of state $\mathbf q$ against itself is lower than that of state $\mathbf p^*$ against $\mathbf q$, which further guarantees the stability of the ESS $\mathbf p^*$. From such a definition, we can see that even if a small fraction of players may not be rational and take out-of-equilibrium strategies, ESS is still a locally stable state. How to find the ESSs is an important issue in EGT. One common approach is to find the stable points of the system state dynamic $\dot{\mathbf p}={d\mathbf p}/{dt}$ \cite{evol2}, i.e, \begin{equation}
\mathbf p^*=\mathop{\arg}\limits_{\mathbf p}\left(\dot{\mathbf p}=0\right).
\end{equation}

The classical evolutionary game theory considers a population in a complete graph. However, in many scenarios, players' spatial locations may lead to an incomplete graph structure. Graphical evolutionary game theory is introduced to study the strategies evolution in such a finite structured population \cite{gegreview}. In graphical EGT, the ``fitness'' of a player is locally determined from interactions with all adjacent players, which is defined as \cite{fitness}
\begin{equation}
\pi=(1-\alpha)\cdot B+\alpha\cdot U,\label{fitness}
\end{equation}
where $B$ is the baseline fitness, which represents the player's inherent property. For example, in a social network, a user's baseline fitness can be interpreted as his/her own interests on the released news. $U$ is the player's payoff which is determined by the predefined payoff matrix and the graph structure. The parameter $\alpha$ represents the selection intensity, i.e., the relative contribution of the game to fitness. The case $\alpha\rightarrow 0$ represents the limit of weak selection \cite{weak}, while $\alpha=1$ denotes strong selection, where fitness equals payoff. The ESS of graphical EGT is also of importance, which is usually analyzed under different strategy updating rules, including birth-death (BD), death-birth (DB) and imitation (IM) introduced as follows \cite{rd}.
\begin{itemize}
\item
BD update rule: a player is chosen for reproduction with the probability being proportional to fitness (Birth process). Then, the chosen player's strategy replaces one neighbor's strategy with uniform probability (Death process), as shown in Fig.\,\ref{fig3}-(a).
\item
DB update rule: a random player is chosen to abandon his/her current strategy (Death process). Then, the chosen player adopts one of his/her neighbors' strategies with the probability being proportional to their fitness (Birth process), as shown in Fig.\,\ref{fig3}-(b).
\item
IM update rule: each player either adopts the strategy of one neighbor or remains with his/her current strategy, with the probability being proportional to fitness, as shown in Fig.\,\ref{fig3}-(c).
\end{itemize}
These strategy update rules are also from the evolutionary biology filed, which are used to model the resident/mutant evolution process. Through analyzing the system dynamics under some specific strategy update rule, the final ESSs can be found at the stable points of the dynamics. In summary, graph structure, players, strategy, fitness (payoff) and ESS are five basic elements of a graphical evolutionary game. In the following, we will illustrate how to formulate the dynamic information diffusion process in a social network using graphical EGT.

\subsection{Graphical Evolutionary Game Formulation}

A social network is usually illustrated by a graph, where each node represents a user and the edge represents the relationship between users. The users can be either human in a social network or websites on the internet, while the relationship can be either friendship between users or hyperlink between webpages. When some new information is originated from one user, the information may be propagated over the network depending on other users' actions: to forward the information or not. For each user, whether to forward the information is determined by several factors, including the user's own interest on this information and his/her neighbor's actions in the sense that if all his/her neighbors forward the information, the user may also forward the information with relatively high probability. In such a case, the users' actions are coupled with each other through their social interactions. This is very similar to the player's strategy update in the aforementioned graphical EGT, where players' strategies are also influenced with each other through the graph structure.

Compared with the strategy updating in the graphical EGT, the information diffusion over social network shares fundamental similarities. A user in a social network can be regarded as a player in a graphical evolutionary game. And each user's two possible actions, i.e., to forward or not forward, are corresponding with two strategies
\begin{eqnarray}
\left\{\begin{array}{ll}
\bm S_f,&\mbox{forward the information}, \vspace{2mm}\\
\bm S_n,&\mbox{not forward the information}.
\end{array}\right.\label{strategy}
\end{eqnarray}
Meanwhile, users' payoff matrix can be written as
\begin{eqnarray}
    \begin{tabular}{ccccc}
    && $\bm S_f$ &$\bm S_n$&\vspace{1mm}\\
    \!\!\!\!$\bm S_f$ \!\!\!\!\!\!\!\!& \multirow{2}{0.01cm}{\bigg(}& $u_{ff}$&$u_{fn}$&\!\!\!\!\!\!\!\!\!\!\multirow{2}{0.01mm}{\bigg)\!\!\!\!}\\
    \!\!\!\!$\bm S_n$ \!\!\!\!\!\!\!\!& & $u_{fn}$&$u_{nn}$&
    \end{tabular}\label{payoff}
\end{eqnarray}
where a symmetric payoff structure is considered, i.e., when a user with strategy $\bm S_f$ meets a user with strategy $\bm S_n$, each of them receives the same payoff $u_{fn}$. Moreover, we assume that the payoff has been normalized within interval $(0,1)$, i.e., $0 < u_{ff},u_{fn},u_{nn} <1$. The physical meaning of the payoff can be either the popularity of a user in a social network or the hit rate of a website. Note that under different application scenarios, the values of the payoff matrix may be different. For example, if the information is related to recent hot topics and forwarding of the information can attract more attentions from other users or website, the payoff matrix should have the following characteristic: $u_{ff}\ge u_{fn}\ge u_{nn}$. On the other hand, if the information is about useless advertisements, the payoff matrix would exhibit $u_{nn}\ge u_{fn}\ge u_{ff}$. Furthermore, if the information is  supposed to be shared only within a circle, i.e., a small group with same interest, the payoff matrix could exhibit $u_{fn}\ge u_{ff}\ge u_{nn}$. It is also worth mentioning that the payoff matrix defined in (\ref{payoff}) is similar to that of the coordination game, where two users making the same actions can obtain symmetric higher payoff than making different actions \cite{corrdination}.

\begin{table}[!t]\renewcommand{\arraystretch}{2}
    \small
    \caption{Correspondence Between Graphical EGT and Social Network.}\label{table3}\vspace{2mm}
    \begin{center}
    \begin{tabular}{|c|c|}
    \hline
    \textbf{Graphical EGT} & \textbf{Social Network}\\ \hline
    Graph structure & Social network topology\\ \hline
    Players & Users in the social network\\ \hline
    \multirow{2}{1cm}{Strategy}& $S_f$: forward the information\\
    & $S_n$: not forward the information\\ \hline
    Fitness & Utility from forwarding or not \\ \hline
    ESS& Stable information diffusion state\\ \hline
    \end{tabular}
    \end{center}
\end{table}

\begin{figure*}[!t]
  \centerline{\epsfig{figure=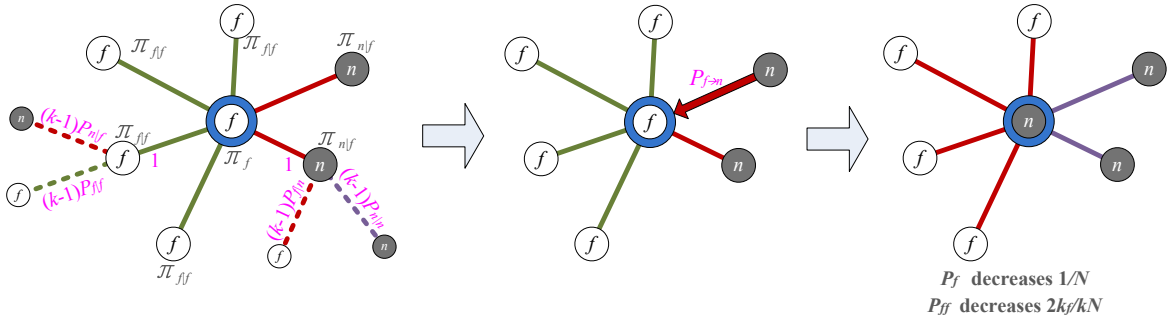,width=15.5cm}}
  \caption{Evolution of user with strategy $\bm S_f$, where $f$ means using strategy $\bm S_f$ and $n$ means using strategy $\bm S_n$.}\label{fig4}
\end{figure*}
Based on the above definitions of strategy and payoff function, we can further define the fitness of users with strategy $\bm S_f$ and strategy $\bm S_n$, respectively. According to (\ref{fitness}), for a user with strategy $\bm S_f$ and $k$ neighbors, among which $k_{f}$ users adopting the same strategy $\bm S_f$ and $k-k_{f}$ users adopting strategy $\bm S_n$, his/her fitness can be written as
\begin{equation}
\pi_f(k,k_{f})=(1-\alpha)+\alpha\left[k_{f} u_{ff}+(k-k_{f})u_{fn}\right],\label{forwardfit}
\end{equation}
where the baseline fitness is normalized as 1. Similarly, for a user with strategy $\bm S_n$ and $k$ neighbors, among which $k_{n}$ users adopting the same strategy $\bm S_n$ and $k-k_{n}$ users adopting strategy $\bm S_f$, his/her fitness can be written as
\begin{equation}
\pi_n(k,k_{n})=(1-\alpha)+\alpha\left[k_{n} u_{nn}+(k-k_{n}) u_{fn}\right].\label{noforwardfit}
\end{equation}

Table~\ref{table3} summarizes the correspondence between the terminologies in graphical EGT and those in the social network. Based on the definitions of strategy and fitness, we can analyze the stable information diffusion state in a social network using graphical EGT. In the following two sections, we will study two scenarios: information diffusion over uniform degree network and non-uniform degree network, respectively.
\begin{figure*}[!t]
  \centerline{\epsfig{figure=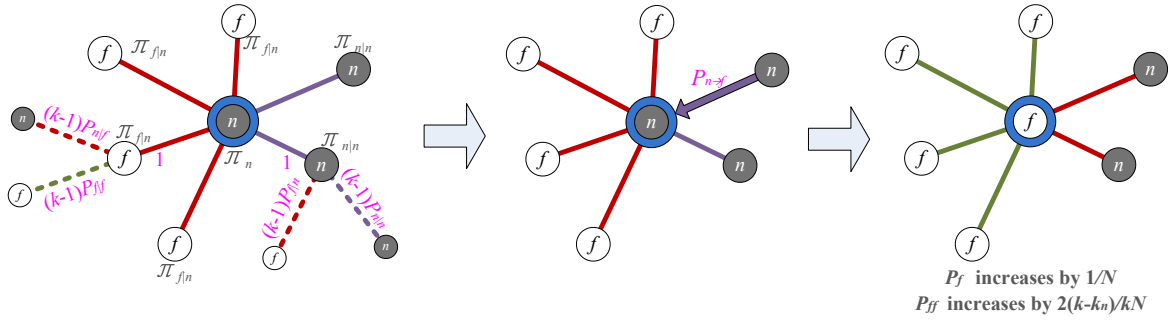,width=15.5cm}}
  \caption{Evolution of user with strategy $\bm S_n$, where $f$ means using strategy $\bm S_f$ and $n$ means using strategy $\bm S_n$.}\label{fig41}
\end{figure*}

\section{Information Diffusion over Uniform Degree Network}

In this section, we study the information diffusion process over networks with uniform degree using graphical EGT. In the uniform scenario, an $N$-user social network based on a homogenous graph with general degree $k$ is considered. When there is new information originated from some user, other users updates their forward strategies, i.e. either forward the information or not ($\bm S_f$ or $\bm S_n$). As for the users' strategy update rule, we model it using the aforementioned IM update rule. Note that all the analysis in the section can be easily extended to the BD and DB update rules. Let us define the network user state as $(p_f,p_n)$, which denotes the percentages of users choosing strategies $\bm S_f$ and $\bm S_n$ among the whole population, respectively. Our ultimate goal is to derive the evolutionary stable network user state $(p^*_f,p^*_n)$, which represents the final diffusion status of the information.

In order to illustrate the network state evolving process on a graph, the edge information of the graph is also required. Let us further define the network edge state as $(p_{ff}, p_{fn}, p_{nn})$, where $p_{ff}$ and $p_{nn}$ represent the percentages of edges on which both users choose same strategy $\bm S_f$ and $\bm S_n$, respectively; $p_{fn}$ represents the percentage of edges on which one user chooses to forward the information while the other chooses not. Moreover, let $p_{f|n}$ denote the percentage of a user's neighbors using strategy $\bm S_f$, given the user is using strategy $\bm S_n$. Similarly, we have $p_{f|f}$, $p_{n|f}$ and $p_{n|n}$. In such a case, according to the basic probability theory, we can summarize the relationship of these defined variables as follows:
\begin{gather}
p_f+p_n=1,\label{eqset1}\\
p_{f|f}+p_{n|f}=1,\ p_{f|n}+p_{n|n}=1,\ p_f p_{n|f}=p_n p_{f|n},\\
\!\!\!p_{ff}=p_f p_{f|f},\ p_{fn}=p_f p_{n|f}+p_n p_{f|n},\ p_{nn}=p_n p_{n|n}.\label{eqset4}
\end{gather}
Eqns. (\ref{eqset1}) and (\ref{eqset4}) imply that the state of the whole social network can be described by only two variables, $p_f$ and $p_{ff}$.
In the following, we will first analyze the dynamics of $p_f$ and $p_{ff}$. Based on the dynamics analysis, the evolutionary stable network state can be achieved.

\subsection{Evolution Dynamics of Network State}

In order to derive the network ESS, we need to analyze the evolution dynamics of the network state. As discussed above, the network state can be represented by parameters $p_f$ and $p_{ff}$, i.e., the network user state and network edge state. In such a case, we will first derive the evolution dynamics of $p_f$ and $p_{ff}$ under the IM update rule, to understand the information diffusion process over the social network. Then, through analyzing the stable point of the network dynamics $\dot p_f$ and $\dot p_{ff}$, the final network ESS can be achieved.

According to the IM update rule, a user adopting strategy $S_f$, i.e., forward the information, is selected for imitation with probability $p_f$. Fig.\,\ref{fig4} illustrates a scenario where the center user adopts strategy $\bm S_f$, and among his/her $k$ neighbors, there are $k_f$ users adopting the same strategy $\bm S_f$ and $k-k_f$ users adopting the opposite strategy $\bm S_n$. The probability of such a configuration is
\begin{equation}
\theta_f(k,k_f)=\binom{k}{k_f}p_{f|f}^{k_f}\big(1-p_{f|f}\big)^{k-k_f}.\label{rho}
\end{equation}
According to the fitness definition in (\ref{forwardfit}), the fitness of the center user in the left part of Fig.\,\ref{fig4}, who adopts strategy $\bm S_f$, can be calculated by
\begin{equation}
\pi_f=(1-\alpha)+\alpha\left[k_{f} u_{ff}+(k-k_{f}) u_{fn}\right].\label{pif}
\end{equation}
Among the center user's $k$ neighbors, for the neighbors adopting strategy $\bm S_f$, each of them has $(k-1)p_{n|f}$ neighbors using strategy $\bm S_n$ and $(k-1)p_{f|f}+1$ neighbors using strategy $\bm S_f$ averagely, where ``1'' means the center user with strategy $\bm S_f$, as shown by the dashed lines in the left part of Fig.\,\ref{fig4}. In such a case, the average fitness of the center user's neighbors who adopt strategy $\bm S_f$, i.e., forward the information, is
\begin{equation}
\pi_{f|f}=(1-\alpha)+\alpha\Big((k-1)p_{n|f} u_{fn}+\big[(k-1)p_{f|f}+1\big] u_{ff}\Big).
\end{equation}
Similarly, for the center user's neighbors adopting strategy $\bm S_n$, each of them has $(k-1)p_{n|n}$ neighbors using strategy $\bm S_n$ and $(k-1)p_{f|n}+1$ neighbors using strategy $\bm S_f$. Thus, the average fitness of the center user's neighbors who adopt strategy $\bm S_n$, i.e., not forward the information, is
\begin{equation}
\pi_{n|f}=(1-\alpha)+\alpha\Big((k-1)p_{n|n} u_{nn}+\big[(k-1)p_{f|n}+1\big] u_{fn}\Big).
\end{equation}

According to the IM update rule, the probability of a user updating his/her strategy is proportional to the fitness of the corresponding strategy.
In such a case, as shown in the middle part of Fig.\,\ref{fig4}, the probability that the center user updates his/her current strategy $\bm S_f$ by strategy $\bm S_n$ is
\begin{equation}
P_{f\rightarrow n}(k_f)=\frac{(k-k_f)\pi_{n|f}}{k_f\pi_{f|f}+(k-k_f)\pi_{n|f}+\pi_f}.
\end{equation}
Therefore, as shown in right part of Fig.\,\ref{fig4}, the center user deviating from using strategy $\bm S_f$ leads to the percentage of users adopting strategy $\bm S_f$, $p_f$, decreasing by $1/N$ with probability
\begin{eqnarray}
\mbox{Prob}\Big(\Delta p_f=-\frac{1}{N}\Big)\!\!\!&=&\!\!\!\sum_{k_f=0}^kp_f\theta_f(k,k_f)P_{f\rightarrow n}(k_f)\nonumber\\
\!\!\!&=&\!\!\!p_f\sum_{k_f=0}^k\binom{k}{k_f}p_{f|f}^{k_f}(1-p_{f|f})^{k-k_f}\nonumber\\
\!\!\!&&\!\!\!\cdot\frac{(k-k_f)\pi_{n|f}}{k_f\pi_{f|f}+(k-k_f)\pi_{n|f}+\pi_f}.\label{pfdec}
\end{eqnarray}
Meanwhile, the center user deviating from using strategy $\bm S_f$ also leads to the edges on which both users adopting strategy $\bm S_f$ decreasing by $k_f$, which means that $p_{ff}$ decreases by $\frac{2k_f}{kN}$ with probability
\begin{eqnarray}
\!\!\!\!\!\!\mbox{Prob}\Big(\Delta p_{ff}=\frac{-2k_f}{kN}\Big)\!\!\!&=&\!\!\!p_f\binom{k}{k_f}p_{f|f}^{k_f}(1-p_{f|f})^{k-k_f}\nonumber\\
\!\!\!&&\!\!\!\cdot\frac{(k-k_f)\pi_{n|f}}{k_f\pi_{f|f}+(k-k_f)\pi_{n|f}+\pi_f}.\label{pffdec}
\end{eqnarray}

Similar analysis can be applied to the user adopting strategy $\bm S_n$. According to the IM update rule, a user adopting strategy $\bm S_n$ is selected for imitation with probability $p_n=1-p_f$. As shown in Fig.\,\ref{fig41}, among its $k$ neighbors, suppose that there are $k_n$ users adopting the same strategy $\bm S_n$ and $k-k_n$ users adopting the opposite strategy $\bm S_f$. The probability of such a configuration is
\begin{equation}
\theta_n(k,k_n)=\binom{k}{k_n} p_{n|n}^{k_n}(1-p_{n|n})^{k-k_n}.\label{rho2}
\end{equation}
Under such a configuration, the fitness of the center user who adopts strategy $\bm S_n$ is
\begin{equation}
\pi_n=(1-\alpha)+\alpha\left[k_n u_{nn}+ (k-k_{n})u_{fn}\right].\label{pin}
\end{equation}
Among his/her $k$ neighbors, the average fitness of neighbors who adopt strategy $\bm S_f$ is
\begin{equation}
\pi_{f|n}=(1-\alpha)+\alpha\Big((k-1)p_{f|f} u_{ff}+\big[(k-1)p_{n|f}+1\big] u_{fn}\Big),
\end{equation}
and the fitness of node adopting strategy $\bm S_n$ is
\begin{equation}
\pi_{n|n}=(1-\alpha)+\alpha\Big((k-1)p_{f|n} u_{fn}+\big[(k-1)p_{n|n}+1\big] u_{nn}\Big),
\end{equation}
In such a case, as shown in the middle part of Fig.\,\ref{fig41}, the probability that the center user updates his/her strategy $\bm S_n$ by strategy $\bm S_f$ is
\begin{equation}
P_{n\rightarrow f}(k_n)=\frac{(k-k_n)\pi_{f|n}}{k_n\pi_{n|n}+(k-k_n)\pi_{f|n}+\pi_n}.
\end{equation}
Therefore, as shown in right part of Fig.\,\ref{fig41}, the center user deviating from using strategy $\bm S_n$ leads to the percentage of users adopting strategy $\bm S_f$, $p_f$, increasing by $1/N$ with probability
\begin{eqnarray}
\mbox{Prob}\Big(\Delta p_f=\frac{1}{N}\Big)\!\!\!&=&\!\!\!\sum_{k_n=0}^k(1-p_f)\theta_n(k,k_n)P_{n\rightarrow f}(k_n)\nonumber\\
\!\!\!&=&\!\!\!(1-p_f)\sum_{k_n=0}^k\binom{k}{k_n} p_{n|n}^{k_n}(1-p_{n|n})^{k-k_n}\nonumber\\
\!\!\!&&\!\!\!\cdot\frac{(k-k_n)\pi_{f|n}}{k_n\pi_{n|n}+(k-k_n)\pi_{f|n}+\pi_n}.\label{pfinc}
\end{eqnarray}
Meanwhile, the deviation also leads to the edges on which both users adopting strategy $\bm S_f$ increasing by $k-k_n$, thus, we have
\begin{eqnarray}
\mbox{Prob}\Big(\Delta p_{ff}=\frac{k-k_n}{kN/2}\Big)\!\!\!\!\!&=&\!\!\!\!\!(1-p_f)\binom{k}{k_n} p_{n|n}^{k_n}(1-p_{n|n})^{k-k_n}\nonumber\\
\!\!\!\!\!&&\!\!\!\!\!\cdot\frac{(k-k_n)\pi_{f|n}}{k_n\pi_{n|n}+(k-k_n)\pi_{f|n}+\pi_n}.\label{pffinc}
\end{eqnarray}

Combining (\ref{pfdec}) and (\ref{pfinc}), we have the probabilities of $p_f$ increasing $1/N$ and decreasing $1/N$, respectively. In such a case, we can calculate the expected variation of $p_f$ per unit time, which is defined as the dynamic of $p_f$, as
\begin{align}
\dot p_f&=\frac{1}{N}\mbox{Prob}\Big(\Delta p_f=\frac{1}{N}\Big)-\frac{1}{N}\mbox{Prob}\Big(\Delta p_f=-\frac{1}{N}\Big)\nonumber\\
&=\frac{\alpha k(k-1)p_{fn}}{2N(k+1)^2}(\gamma_1u_{nn}+\gamma_2u_{fn}+\gamma_3u_{ff})+O(\alpha^2),\label{dpm}
\end{align}
where the second equality is according to Taylor's Theorem and weak selection assumption with $\alpha$ goes to zero \cite{weak0}, and the parameters $\gamma_1$, $\gamma_2$ and $\gamma_3$ are given as follows:
\begin{align}
\gamma_1=&\ -p_{n|n}[(k-1)(p_{n|n}+p_{f|f})+3],\label{steady}\\
\gamma_2=&\ p_{n|n}-p_{f|f}+(p_{n|f}-p_{f|n})\nonumber\\
&\ \cdot[(k-1)(p_{n|n}+p_{f|f})+2],\\
\gamma_3=&\ p_{f|f}[(k-1)(p_{n|n}+p_{f|f})+3]\label{steady2}.
\end{align}
Under the weak selection, the payoff obtained from the interactions is considered as limited contribution to the overall fitness of each player. The results derived from weak selection often remain as valid approximations for larger selection strength \cite{weak}. Moreover, the weak selection has a long tradition in theoretical biology \cite{weak1}. Moreover, the weak selection assumption can help achieve a close-form analysis of diffusion process and better reveal how the strategy diffuses over the network. From (\ref{dpm}), we can see that the dynamics $\dot p_f$ is an increasing function in terms of $u_{ff}$, i.e., the payoff of both users forwarding the released information. The corresponding physical meaning is that when the higher payoff can be obtained by forwarding the information, the increasing rate of $p_f$ will also be higher. On the other hand, if not forwarding the information can gain higher payoff, the increasing rate of $p_f$ will be lower, which is just the reason why $\dot p_f$ is a decreasing function in terms of $u_{nn}$, i.e., the payoff of both users do not forward the information.
Similarly, by combining (\ref{pffdec}) and (\ref{pffinc}), we have the probabilities of $p_{ff}$ increasing $2k_f/kN$ and decreasing $2(k-k_n)/kN$, respectively. Thus, we can calculate the expected variation of $p_{ff}$ per unit time, which is defined as the dynamic of $p_{ff}$, as
\begin{align}
\dot p_{ff}=&\ \sum\limits_{k_f=0}^k\frac{-2k_f}{kN}\mbox{Prob}\Big(\Delta p_{ff}=\frac{-2k_f}{kN}\Big)\nonumber\\
&\ +\sum\limits_{k_n=0}^k\frac{2(k-k_n)}{kN}\mbox{Prob}\Big(\Delta p_{ff}=\frac{2(k-k_n)}{kN}\Big)\nonumber\\
=&\ \frac{p_{fn}}{(k+1)N}\Big(1+(k-1)(p_{f|n}-p_{f|f})\Big)+O(\alpha).\label{dpff}
\end{align}

\subsection{ESS of Information Diffusion}

As discussed in Section II-A, at the ESS, the network state dynamics achieve stable point, i.e., $\dot p_f=0$ and $\dot p_{ff}=0$. By setting $\dot p_{ff}=0$, we have
\begin{equation}
p_{f|f}-p_{f|n}=\frac{1}{k-1}.\label{steady3}
\end{equation}
In such a case, according to (\ref{eqset1})-(\ref{eqset4}) and (\ref{steady3}), all the network state parameters can be expressed only by $p_f$ as follow:
\begin{eqnarray}
p_{f|f}\!\!\!&=&\!\!\!p_f+\frac{1}{k-1}(1-p_f),\label{pffpf}\\
p_{f|n}\!\!\!&=&\!\!\!\frac{k-2}{k-1}p_f,\\
p_{n|f}\!\!\!&=&\!\!\!\frac{k-2}{k-1}(1-p_f),\\
p_{n|n}\!\!\!&=&\!\!\!1-\frac{k-2}{k-1}p_f\label{pnnpf}.
\end{eqnarray}
Then, by substituting (\ref{pffpf})-(\ref{pnnpf}) and (\ref{steady})-(\ref{steady2}) into (\ref{dpm}), we can obtain the dynamic of the network user state, $\dot p_f$, as
\begin{eqnarray}
\dot p_f=\frac{\alpha k(k-2)(k+3)}{2N(k-1)(k+1)^2}p_f(1-p_f)(ap_f-b),\label{dynamic}
\end{eqnarray}
where the parameters $a$ and $b$ are given as follows:
\begin{align}
a&=(k-2)(u_{ff}-2u_{fn}+u_{nn}),\label{a}\\
b&=(k-1)u_{nn}-(k-2)u_{fn}-u_{ff}.\label{b}
\end{align}
At the ESS, we have $\dot p_f=0$. Therefore, there are three evolutionary stable network states:
\begin{equation}
p^*_f=0, 1 \mbox{ and } \frac{b}{a}.
\end{equation}
The following theorem shows the conditions of these stable network states.

\emph{Theorem 1:} In an $N$-user social network which can be characterized by a graph with uniform degree $k$, if each user updates his/her information forward strategy using the IM update rule, the evolutionary stable network states can be summarized as follows:
\begin{equation}
p^*_f=\left\{
\begin{array}{lc}
1,& \mbox{if }u_{ff}>u_{fn}>u_{nn},\\
0,& \mbox{if }u_{nn}>u_{fn}>u_{ff},\\
\frac{(k-2)u_{fn}+u_{ff}-(k-1)u_{nn}}{2u_{fn}-(k-2)(u_{ff}-u_{nn})},&\mbox{else}.
\end{array}
\right.\label{udpf}
\end{equation}
\begin{proof}
Considering (\ref{dpm}) and (\ref{dpff}) as a nonlinear dynamic system, we can examine whether the three stable points are evolutionary stable network states through analyzing the Jacobian matrix of the dynamics as follows:
\begin{equation}
\mathbf J=\begin{bmatrix}{\partial{\dot{p}_f}}/{\partial{p_f}}&{\partial{\dot{p}_f}}/{\partial{p_{ff}}}\\
{\partial{\dot{p}_{ff}}}/{\partial{p_f}}&{\partial{\dot{p}_{ff}}}/{\partial{p_{ff}}}\end{bmatrix}.\label{jacob}
\end{equation}
The evolutionary stability requires that $det(\mathbf J)>0$ and $tr(\mathbf J)<0$ \cite{evol}. By substituting (\ref{dpm}) and (\ref{dpff}) into (\ref{jacob}), we can obtain the three conditions for $p^*_f=0, 1 \mbox{ and } \frac{b}{a}$, respectively. Due to page limitation, the detailed derivations are omitted here.
\end{proof}

Among these three ESSs, $p^*_f=0$ and $1$ are two extreme stable states, representing no user forwarding the information and all users forwarding the information, respectively. According to \emph{Theorem 1}, we can see that the condition of $p^*_f=1$ means that both users forwarding the information can gain the most payoff, while not forwarding gains the least payoff. In a social network, this is corresponding to the scenario where the released information is an extremely hot topic, forwarding which can attract more attentions. On the contrary, the condition of $p^*_f=0$ means that neither user forwarding the information can gain the most payoff, while both forwarding gains the least payoff, which is corresponding to the scenario where the released information is useless or negative advertisement, forwarding which can only incur unnecessary cost. For the third ESS, $p^*_f=\frac{b}{a}$, some approximations can be made as follows:
\begin{align}
p^*_f=\frac{b}{a}&=\frac{(k-2)u_{fn}+u_{ff}-(k-1)u_{nn}}{(k-2)(2u_{fn}-u_{ff}-u_{nn})}\nonumber\\
&=\frac{(k-1)(u_{fn}-u_{nn})+(u_{ff}-u_{fn})}{(k-2)(u_{fn}-u_{ff}+u_{fn}-u_{nn})}\nonumber\\
&\ \dot=\frac{1}{1+\frac{u_{fn}-u_{ff}}{u_{fn}-u_{nn}}},\label{appro}
\end{align}
where the last approximation is based on the assumption that the network degree $k\gg 2$ in real social networks. In such a case, we can further discuss some physical meanings of the third ESS as follows.

\begin{itemize}
\item \emph{Case 1:} $u_{fn}>u_{ff}$ and $u_{fn}>u_{nn}$.
\end{itemize}

In this case, unilateral forwarding can bring more payoff than no forwarding or both forwarding. In a social network, this case is corresponding to the scenario where both users forwarding the information can gain only limited reward but incurring more cost to both of them. An example of this case can be that the information is not the mainstream topic, e.g. the news about a punk musician, and is supposed to be diffused among people with similar interstates. Specifically, when $u_{ff}=u_{nn}$, we have $p^*_f\approx 0.5$. This means that if there is no payoff difference between forwarding and not forwarding the information, the final stable network state tends to ``half and half''. Moreover, when $u_{fn}>u_{ff}>u_{nn}$, we have $p^*_f> 0.5$, i.e., more than half of users' strategies are forwarding; while when $u_{fn}>u_{nn}>u_{ff}$, we have $p^*_f< 0.5$, i.e., less than half of users' forward the information.

\begin{itemize}
\item \emph{Case 2:} $u_{ff}>u_{fn}$ and $u_{nn}>u_{fn}$.
\end{itemize}

This is the opposite case of \emph{Case 1}. In this case, the payoff configuration is equivalent to that of the coordination game, where both players with the same actions can make more payoff than opposite actions. In a social network, this case is corresponding to the scenario where both users forwarding the information can enhance the popularity of them simultaneously with some forwarding costs. However, if only one user forwards the information, the associated cost is larger than the unilateral benefit. An example of this case can be that the information is politically sensitive and the reality of it is not guaranteed, forwarding which may gain attractions but also incurring potential misleading cost.

\section{Information Diffusion over non-Uniform Degree Network}

In this section, we study the information diffusion process over networks with non-uniform degree. In the non-uniform scenario, we consider an $N$-user social network based on a graph whose degree exhibits distribution $\lambda(k)$. This distribution means that when randomly choosing one user on the network, the probability of the chosen user with $k$ neighbors is $\lambda(k)$. In such a case, the average degree of the network is
\begin{equation}
\overline k=\sum_{k=0}^{+\infty}\lambda(k)k.\label{ak}
\end{equation}
Note that we do not take degree correlation into account, i.e., the degrees of all users are independent of each other. Similar to the analysis in Section III, when some new information is released, all users update their information forwarding strategies in a spontaneous manner. In order to give more insights on different evolutionary update rules, in this section, we model the users' information forwarding strategy update using the aforementioned BD update rule. In the following, we first analyze the general case of information diffusion over non-uniform degree networks. Then, we focus on two special cases, i.e., two typical random networks, Erd\H{o}s-R\'enyi random network and Barab\'asi-Albert scale-free networks.

\subsection{General Case}

In the analysis of non-uniform degree network, to ensure the derivation of ESS is mathematically realizable, we assume that the network user state $p_f$ and the network edge states $p_{ff}$, $p_{fn}$ and $p_{nn}$, do not depend on the degree.  With such an assumption, eqns. (\ref{eqset1}) and (\ref{eqset4}) still hold for the non-uniform degree network, i.e., the network state can still be described by two variables $p_f$ and $p_{ff}$. As a matter of fact, it has been shown that the dependence of the network state on degree is weak \cite{dependence}. Similar to the analysis in Section III, in order to find the evolutionary stable network state, we need to first derive the evolving dynamics of $p_f$ and $p_{ff}$.

According to the BD update rule, a user is chosen with the probability being proportional to fitness. Then, the chosen user's strategy replaces one of his/her neighbors' strategies uniformly. As shown in Fig.\,\ref{fig4}, suppose the chosen user is adopting strategy $\bm S_f$ and with degree $l$. Among his/her neighbors, there are $l_f$ users adopting the same strategy $\bm S_f$ and $l-l_f$ users adopting the opposite strategy $\bm S_n$. In such a case, the probability that the chosen user's strategy $\bm S_f$ replaces one of his/her neighbors adopting strategy $\bm S_n$ is
\begin{equation}
P_{n\rightarrow f}(l,l_f)=p_f\theta_f(l,l_f)\frac{\pi_f(l,l_f)}{\overline \pi}\frac{l-l_f}{l},
\end{equation}
where $\theta_f(l,l_f)$ is the probability of the configuration given in (\ref{rho}), $\pi_f(l,l_f)$ is the fitness of the chosen user given in (\ref{forwardfit}) and $\overline\pi$ is the expected fitness of all users which can be calculated as follows:
\begin{align}
\overline \pi=&\ Np_f\sum_{k=0}^{+\infty}\lambda(k)\big(kp_{f|f}u_{ff}+k(1-p_{f|f})u_{fn}\big)+\nonumber\\
&\quad Np_n\sum_{k=0}^{+\infty}\lambda(k)\big(kp_{n|n}u_{nn}+k(1-p_{n|n})u_{fn}\big)\nonumber\\
=&\ N(p_{ff}u_{ff}+p_{fn}u_{fn}+p_{nn}u_{nn}).\label{api}
\end{align}
Therefore, the percentage of users adopting strategy $\bm S_f$, $p_f$, increases by $1/N$ with expected probability
\begin{equation}
\mbox{Prob}\Big(\Delta p_f\!=\!\frac{1}{N}\Big)\!=\!\sum_{l=0}^{+\infty}\lambda(l)\!\sum_{l_f=0}^lp_f\theta_f(l,l_f)\frac{\pi_f(l,l_f)}{\overline \pi}\frac{l-l_f}{l}.\label{npf1}
\end{equation}
Suppose the replaced user, who was adopting strategy $\bm S_n$, is with degree $l^\prime$. After strategy update, the edges on which both users adopting strategy $\bm S_f$ will increase by $(l^\prime-1)p_{f|n}+1$. In such a case, we have
\begin{equation}
\mbox{Prob}\Big(\Delta p_{ff}=\frac{(l^\prime-1)p_{f|n}+1}{\overline kN/2}\Big)=\mbox{Prob}\Big(\Delta p_f=\frac{1}{N}\Big),\!\!\!\label{npff1}
\end{equation}
where $\overline k$ is the average degree of the network given in (\ref{ak}).

Similarly, suppose the chosen user is adopting strategy $\bm S_n$ and with degree $m$ as shown in Fig.\,\ref{fig4}. Among his/her neighbors, there are $m_n$ users adopting the same strategy $\bm S_f$ and $m-m_n$ users adopting the opposite strategy $\bm S_n$. In such a case, according to the BD update rule, the percentage of users adopting strategy $\bm S_f$, $p_f$, decreases by $1/N$ with expected probability
\begin{align}
&\!\!\!\!\!\!\!\!\!\!\mbox{Prob}\Big(\Delta p_f=-\frac{1}{N}\Big)=\sum_{m=0}^{+\infty}\lambda(m)\nonumber\\
&\sum_{m_f=0}^m(1-p_f)\theta_n(m,m_f)\frac{\pi_n(m,m_f)}{\overline \pi}\frac{m_f}{m},\label{npf2}
\end{align}
where $\theta_n(k,k_f)$ is given in (\ref{rho2}), $\pi_n(k,k_f)$ is given in (\ref{noforwardfit}) and $\overline \pi$ is given in (\ref{api}).
Suppose the replaced user, who was adopting strategy $\bm S_f$, is with degree $m^\prime$. After strategy update, the edges on which both users adopting strategy $\bm S_f$ will decrease by $(m^\prime-1)p_{f|f}$. In such a case, we have
\begin{equation}
\mbox{Prob}\Big(\Delta p_{ff}=-\frac{(m^\prime-1)p_{f|f}}{\overline kN/2}\Big)=\mbox{Prob}\Big(\Delta p_f=-\frac{1}{N}\Big).\!\!\!\label{npff2}
\end{equation}

Combining (\ref{npf1}) and (\ref{npf2}), we have the dynamics of $p_f$ as
\begin{align}
&\dot p_f=\frac{1}{N}\mbox{Prob}\Big(\Delta p_f=\frac{1}{N}\Big)-\frac{1}{N}\mbox{Prob}\Big(\Delta p_f=-\frac{1}{N}\Big)=\nonumber\\
&\frac{\alpha(\overline k-1)p_{fn}}{2\overline \pi N}\big[p_{f|f}(u_{ff}-u_{fn})-p_{n|n}(u_{nn}-u_{fn})\big]+O(\alpha^2),\label{50}
\end{align}
where $\overline k$ is the average degree of the network given in (\ref{ak}).
Combining (\ref{npff1}) and (\ref{npff2}), we have the dynamics of $p_{ff}$ as
\begin{align}
\dot p_{ff}=&\ \frac{(\overline {l^\prime}-1)p_{f|n}+1}{\overline kN/2}\cdot\mbox{Prob}\Big(\Delta p_f=\frac{1}{N}\Big)\nonumber\\
&-\frac{(\overline {m^\prime}-1)p_{f|f}}{\overline kN/2}\cdot\mbox{Prob}\Big(\Delta p_f=-\frac{1}{N}\Big),\label{52}
\end{align}
where $\overline {l^\prime}$ and $\overline {m^\prime}$ are the average degrees of the replaced users during the strategy update, which are different with the average degree of the whole network, $\overline k$. If a link is selected at random, the degree distribution of the node on the specific link is not $\lambda(k)$ but rather $\frac{k\lambda(k)}{\sum_{k=0}^{+\infty}k\lambda(k)}$ \cite{degreec}. Thus, the average degree of the replaced users is
\begin{equation}
\overline {l^\prime}=\overline {m^\prime}=\sum_{k=0}^{+\infty}k\frac{k\lambda(k)}{\sum_{k=0}^{+\infty}k\lambda(k)}=\frac{\overline {k^2}}{\overline k},\label{53}
\end{equation}
where $\overline {k^2}=\sum_{k=0}^{+\infty}k^2\lambda(k)$ is the expectoration of $k^2$. Based on the dynamics of $p_f$ and $p_{ff}$ in (\ref{50}) and (\ref{52}), we can obtain the following theorem.

\emph{Theorem 2:} In an $N$-user social network which can be characterized by a graph with degree distribution $\lambda(k)$, if each user updates his/her information forward strategy using the BD update rule, the evolutionary stable network states can be summarized as follows:
\begin{equation}
p^*_f=\left\{
\begin{array}{lc}
1,& \!\!\!\!\mbox{if }u_{ff}>u_{fn}>u_{nn},\\
0,& \!\!\!\!\mbox{if }u_{nn}>u_{fn}>u_{ff},\\
\frac{(\overline {k^2}/\overline k-2)(u_{fn}-u_{nn})+(u_{ff}-u_{nn})}{(\overline {k^2}/\overline k-2)(2u_{fn}-u_{ff}-u_{nn})},&\mbox{else}.
\end{array}\label{theorem2c}
\right.
\end{equation}
\begin{proof}
Similar to the proof of \emph{Theorem 1}, the evolutionary stable network state $p^*_f$ can be derived by solving $\dot p_f=0$ and $\dot p_{ff}=0$ in (\ref{52}) and (\ref{53}). Meanwhile, the conditions in (\ref{theorem2c}) can be obtained by analyzing the Jacob matrix of $\dot p_f$ and $\dot p_{ff}$. Due to page limitation, the detailed derivation is omitted.
\end{proof}
\subsection{Special Cases}

In this subsection, we discuss two special cases of the non-uniform degree networks, Erd\H{o}s-R\'enyi random network \cite{er} and Barab\'asi-Albert scale-free network \cite{ba}. For the Erd\H{o}s-R\'enyi random (ER) network, the degree follows a Poisson distribution, i.e.,
\begin{equation}
\lambda_{\mbox{\scriptsize ER}}(k)=\frac{e^{-\overline k}\overline k^k}{k!}\ \mbox{ and }\ \overline {k^2}=\overline k(\overline k+1).
\end{equation}
In such a case, according to \emph{Theorem 2}, the ESS of Erd\H{o}s-R\'enyi random network is
\begin{equation}
p^*_{f\mbox{\scriptsize ER}}=\left\{
\begin{array}{lc}
1,& \!\!\!\!\mbox{if }u_{ff}>u_{fn}>u_{nn},\\
0,& \!\!\!\!\mbox{if }u_{nn}>u_{fn}>u_{ff},\\
\frac{(\overline k-1)(u_{fn}-u_{nn})+(u_{ff}-u_{nn})}{(\overline k-1)(2u_{fn}-u_{ff}-u_{nn})},&\mbox{else}.
\end{array}
\right.\label{erpf}
\end{equation}
For the Barab\'asi-Albert scale-free (BA) network, the degree follows a power law distribution, i.e.,
\begin{equation}
\lambda_{\mbox{\scriptsize BA}}(k)\varpropto k^{-\xi}\ \mbox{ and }\ \overline {k^2}\dot =\overline k^2\log N/4\ (\mbox{when }\xi=3).
\end{equation}
In such a case, according to \emph{Theorem 2}, the ESS of Barab\'asi-Albert scale-free network is
\begin{equation}
p^*_{f\mbox{\scriptsize BA}}=\left\{
\begin{array}{lc}
1,&\!\!\!\!\!\!\!\!\!\!\!\!\!\!\!\!\! \mbox{if }u_{ff}>u_{fn}>u_{nn},\\
0,&\!\!\!\!\!\!\!\!\!\!\!\!\!\!\!\!\! \mbox{if }u_{nn}>u_{fn}>u_{ff},\\
\frac{(\overline k\log N-8)(u_{fn}-u_{nn})+4(u_{ff}-u_{nn})}{(\overline k\log N-8)(2u_{fn}-u_{ff}-u_{nn})},&\!\!\!\!\!\!\!\!\!\!\!\!\mbox{else}.
\end{array}
\right.\label{bapf}
\end{equation}

\section{Experiments}

In this section, we conduct experiments to verify the information diffusion analysis. First, we simulate the information diffusion process on synthetic networks. Then, we conduct information diffusion experiment on real-world networks, i.e., the Facebook social network. Finally, we use real-world information diffusion dataset to further verify the diffusion analysis.

\subsection{Synthetic Networks}

In the experiment of synthetic networks, we generate three kinds of networks to simulate the information diffusion process:
\begin{itemize}
\item the uniform-degree network;
\item the Erd\H{o}s-R\'enyi random network;
\item the Barab\'asi-Albert scale-free network.
\end{itemize}
For each network, we generate 1000 users and initialize each user with random strategy: $S_f$ or $S_n$. In the simulations, four kinds of payoff matrices are considered as follows:
\begin{itemize}
\item PM 1: $u_{ff}>u_{fn}>u_{nn}$
\begin{eqnarray}
    \begin{tabular}{ccccc}
    && $\bm S_f$ &$\bm S_n$&\vspace{1mm}\\
    \!\!\!\!$\bm S_f$ \!\!\!\!\!\!\!\!& \multirow{2}{0.01cm}{\bigg(}& $0.8$&$0.6$&\!\!\!\!\!\!\!\!\!\!\multirow{2}{0.01mm}{\bigg)\!\!\!\!}\\
    \!\!\!\!$\bm S_n$ \!\!\!\!\!\!\!\!& & $0.6$&$0.4$&
    \end{tabular}
\end{eqnarray}
\item PM 2: $u_{fn}>u_{ff}>u_{nn}$
\begin{eqnarray}
    \begin{tabular}{ccccc}
    && $\bm S_f$ &$\bm S_n$&\vspace{1mm}\\
    \!\!\!\!$\bm S_f$ \!\!\!\!\!\!\!\!& \multirow{2}{0.01cm}{\bigg(}& $0.6$&$0.8$&\!\!\!\!\!\!\!\!\!\!\multirow{2}{0.01mm}{\bigg)\!\!\!\!}\\
    \!\!\!\!$\bm S_n$ \!\!\!\!\!\!\!\!& & $0.8$&$0.4$&
    \end{tabular}
\end{eqnarray}
\item PM 3: $u_{fn}>u_{nn}>u_{ff}$
\begin{eqnarray}
    \begin{tabular}{ccccc}
    && $\bm S_f$ &$\bm S_n$&\vspace{1mm}\\
    \!\!\!\!$\bm S_f$ \!\!\!\!\!\!\!\!& \multirow{2}{0.01cm}{\bigg(}& $0.4$&$0.8$&\!\!\!\!\!\!\!\!\!\!\multirow{2}{0.01mm}{\bigg)\!\!\!\!}\\
    \!\!\!\!$\bm S_n$ \!\!\!\!\!\!\!\!& & $0.8$&$0.6$&
    \end{tabular}
\end{eqnarray}
\item PM 4: $u_{nn}>u_{fn}>u_{ff}$
\begin{eqnarray}
    \begin{tabular}{ccccc}
    && $\bm S_f$ &$\bm S_n$&\vspace{1mm}\\
    \!\!\!\!$\bm S_f$ \!\!\!\!\!\!\!\!& \multirow{2}{0.01cm}{\bigg(}& $0.4$&$0.6$&\!\!\!\!\!\!\!\!\!\!\multirow{2}{0.01mm}{\bigg)\!\!\!\!}\\
    \!\!\!\!$\bm S_n$ \!\!\!\!\!\!\!\!& & $0.6$&$0.8$&
    \end{tabular}
\end{eqnarray}
\end{itemize}

\begin{figure}[!t]
  \centerline{\epsfig{figure=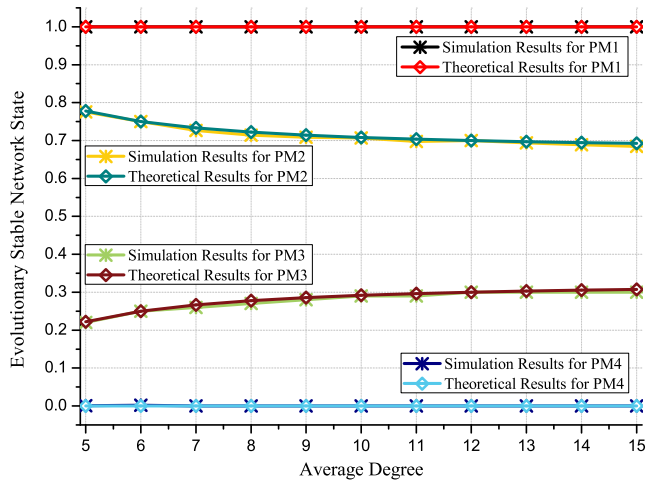,width=8.5cm}}
  \caption{Simulation results for the uniform-degree network.}\label{uni}
\end{figure}

Fig.\,\ref{uni} shows the experiment results for the uniform-degree network under different average degrees and payoff matrices. The theoretical results is calculated from (\ref{udpf}) directly, while the simulation results are obtained by simulating the IM strategy update rule over the generated network. For each simulation run, the strategy update steps are repeated until the network reaches the stable network state. Meanwhile, the network structure is re-generated every $500$ runs to prevent any spurious results based on one particular realization of a specific network type. From Fig.\,\ref{uni}, we can see that all the simulation results are consistent with the theoretical results, which verifies the correctness of our diffusion analysis and the conclusion in \emph{Theorem 1}. Moreover, different settings of the payoff matrix can lead to different evolutionary stable network states, while the network degree variations have little influence on the stable network states. Such a structure-free phenomenon is consistent with our approximation analysis for the large-degree case after \emph{Theorem 1}, especially in (\ref{appro}).

\begin{figure}[!t]
  \centerline{\epsfig{figure=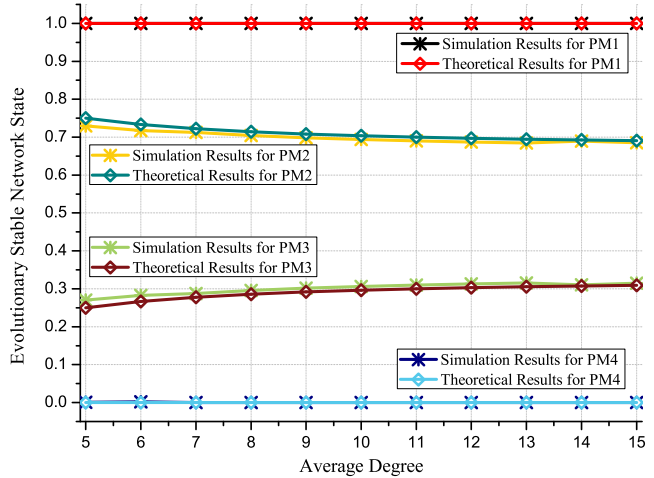,width=8.5cm}}
  \caption{Simulation results for the Erd\H{o}s-R\'enyi random network.}\label{ers}
\end{figure}

\begin{figure}[!t]
  \centerline{\epsfig{figure=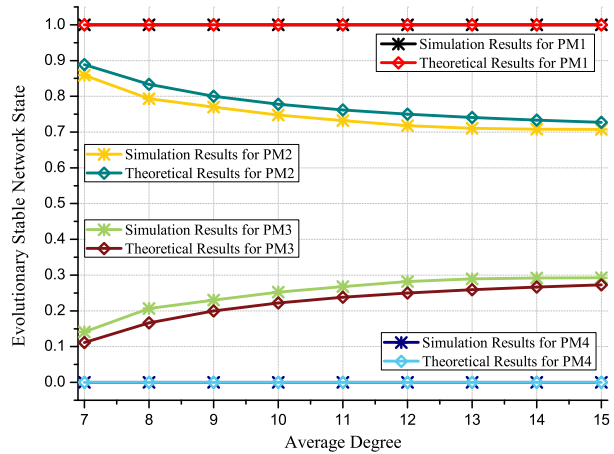,width=8cm}}
  \caption{Simulation results for the Barab\'asi-Albert scale-free network.}\label{abs}
\end{figure}
Fig.\,\ref{ers} and \ref{abs} show the experiment results for the non-uniform degree networks under different average degrees and payoff matrices, including the Erd\H{o}s-R\'enyi random network and the Barab\'asi-Albert scale-free network. The theoretical results are calculated from (\ref{erpf}) and (\ref{bapf}) directly, while the simulation results are obtained by simulating the BD strategy update rule over the two generated networks. We can see that the all the simulation results agree well with the theoretical results. In Fig.\,\ref{abs}, the small gap for the Barab\'asi-Albert scale-free networks is due to the fact that there is weak dependence between the network state and the network degree, while we neglected such dependence in the diffusion analysis. Nevertheless, we can see that the gap is relatively small and indeed negligible.

\begin{figure}[!t]
  \centerline{\epsfig{figure=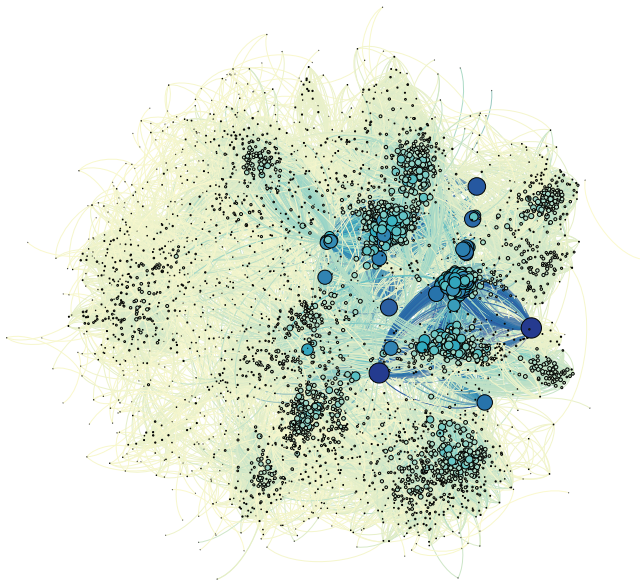,width=8.5cm}}
  \caption{Graph structure of the Facebook network used for experiment.}\label{graph}
\end{figure}
\begin{figure}[!t]
  \centerline{\epsfig{figure=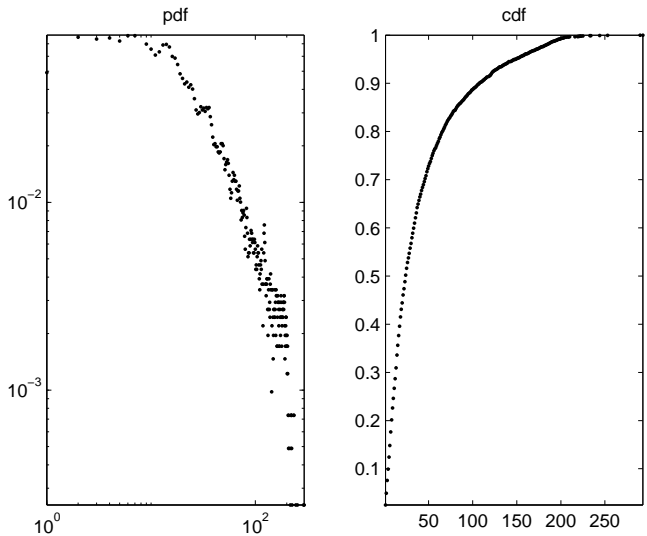,width=8.5cm}}
  \caption{Degree characteristics of the Facebook network.}\label{degree}
\end{figure}

\subsection{Real-World Networks}

In the experiment of real-world network, we evaluate the information diffusion process over Facebook social network \cite{facebookgrpah}. The Facebook dataset contains totally 4039 users and 88234 edges, where the edge means the connection between two users \cite{dataset}. The abstract graph of the Facebook network is depicted in Fig.\,\ref{graph}, where the nodes with large degrees, i.e., the users with large number of friends, are highlighted. Meanwhile, we also plot the degree characteristics of the Facebook network in Fig.\,\ref{degree}, including the cumulative distribution (CDF) and probability density (PDF) of the network users' degrees. From the CDF, we can see that the users' degrees vary from 1 to 300 and 90 percents of the users are with less than 100 degrees. From the PDF where the axes are in logarithm scale, we can see that the Facebook network degree exhibits the scale-free phenomenon \cite{ba}.

\begin{figure}[!t]
  \centerline{\epsfig{figure=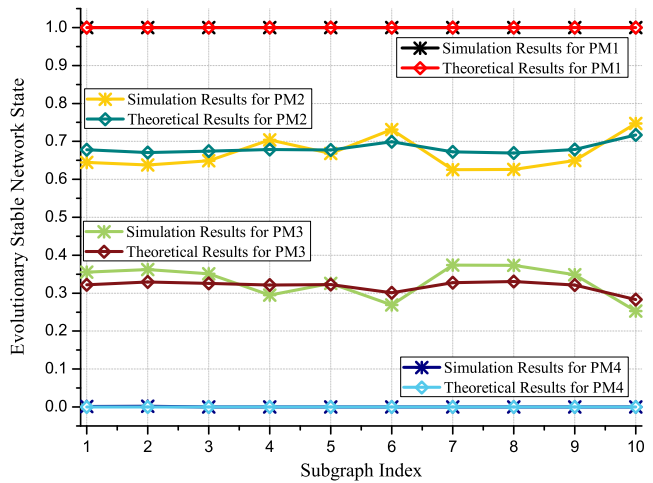,width=8.5cm}}
  \caption{Experiment result for the Facebook network.}\label{facebook}
\end{figure}

The Facebook dataset contains ten ego-networks, which means that there are ten subgraphs. In the experiment, we simulate the information diffusion process over the ten subgraphs respectively. Fig.\,\ref{facebook} shows the experiment results under different payoff matrix settings. The theoretical results are calculated from (\ref{theorem2c}), while the simulation results are obtained by simulating the BD strategy update rule over the ten subgraphs. It can be seen that the simulation results match well with the theoretical results for all ten subgraphs. The small gaps for some figures, e.g., the 10-th subgraph, are mainly due to the neglected dependence between the network state and the network degree. Since the average degrees of all subgraphs are similar with each other, the evolutionary network stable states of all subgraphs are also similar with each other.

\subsection{Information Diffusion Dataset Evaluation}
\begin{figure*}[!t]
  \centerline{\epsfig{figure=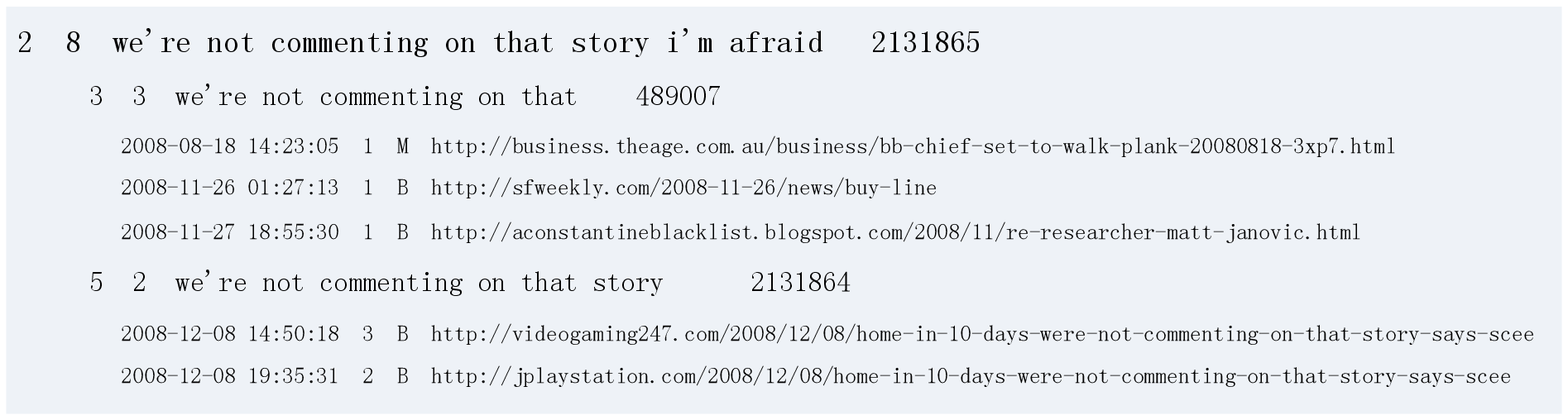,width=16cm}}
  \caption{Example of the Memetracker phrase cluster dataset \cite{meme}.}\label{dataset}
\end{figure*}
In this subsection, we further verify our diffusion analysis using real-world information spreading dataset \cite{meme}. In the previous two subsections, we first set the users' payoff matrix, and then conducted experiments to find the final stable state. In this experiment, an inverse step is conducted, we first estimate the evolutionary stable state through mining the real-world information spreading dataset, and then train the corresponding payoff matrix. The dataset used for experiment, named ``MemeTracker'', contains more than 172 million news articles and blog posts from 1 million online sources \cite{meme2}. When a site publishes a new post, it will put hyperlinks to related posts in some other sites published earlier as its sources. And later, the site will also be cited by other newer posts as well. In such a case, the hyperlinks between articles and posts can be used to represent the spreading of information from one site to another site.

\begin{table}[!t]\renewcommand{\arraystretch}{2}
    \small
    \caption{Experiment results for Memetracker dataset.}\label{table}\vspace{2mm}
    \begin{center}
    \begin{tabular}{|c|c|c|c|c|c|}
    \hline
    \textbf{Group Index} &1 &2 &3 &4 &5\\ \hline
    \textbf{ESS} & 0.19 &0.35 & 0.53 & 0.77 & 0.81\\\hline
    \end{tabular}
    \end{center}
\end{table}

In the MemeTracker phrase cluster dataset, each ``phrase cluster'' contains all the plain and mutant phrases in one cluster, a list of sites where the phrases appeared, and time indexes when the phrase appeared. An example of one phrase cluster in the dataset is shown in Fig.\,\ref{dataset}. In such a case, each phrase cluster can be regarded as a diffusion process of one piece of information. We extract 5 group of sites, where each group includes 500 sites. Each group is regarded as a complete graph and each site is considered as a user in our information diffusion game. We divide the dataset into two halfs, where the first half is used to train the payoff matrix and the second half is used for testing. Through calculating the average hyperlinks of all phrases clusters in each group, we can obtain the statistical evolutionary stable state of each group using the first half dataset, as shown in Table~\ref{table}. We can see that the sites in group 5 share major common interests, while the sites in group 1 share relatively rare common interests. Using our proposed game theoretic analysis and the data-mining based approach, enterprises/polititians can classify users into different same-interest categories according to the stable states of different information diffusions, which can help them to achieve more effective advertisement/advocation.

\begin{figure}[!t]
  \centerline{\epsfig{figure=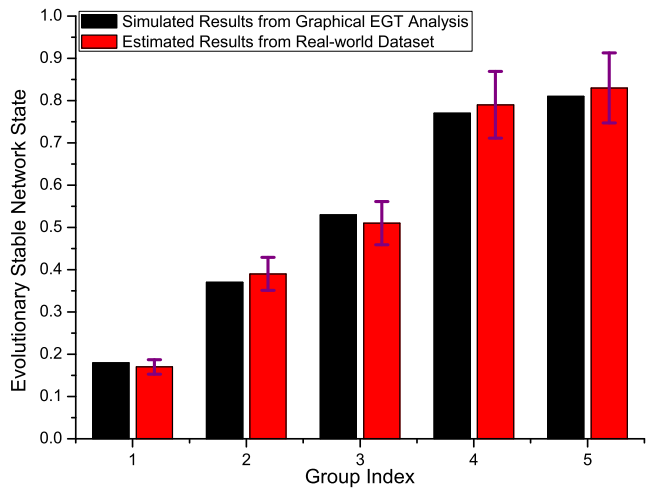,width=8.5cm}}
  \caption{Experiment result for Memetracker dataset.}\label{memediff}
\end{figure}

Since the site network of each group is considered as a uniform degree network, by taking Table~\ref{table} back to \emph{Theorem 1} and normalize $u_{fn}$ as 1, we can find the relationship between $u_{ff}$ and $u_{nn}$. Based on the relationship, we can parameterize the payoff matrix of each group, simulate their information diffusion processes on uniform degree graph and achieve the simulated stable information diffusion state of each group. Meanwhile, we can also test the stable information diffusion state of each group using the second half of the Memetracker dataset. Fig.\,\ref{memediff} shows the simulated results from our game theoretical analysis and the estimated results from the real-world dataset, from which we can see they match well with each other. We also depict the variances of the estimated results in Fig.\,\ref{memediff}, which shows that the simulated results are always in the variance interval of the corresponding estimated results. This further verifies that our graphical evolutionary game theoretical analysis for the information diffusion over social network is effective and practical.

%

\section{Conclusion}

In this paper, we formulate the information diffusion problem using graphical evolutionary game theory. We defined the players, strategies and payoff matrix in this problem, and highlight the correspondence between graphical EGT and information diffusion. Two kinds of networks, uniform and non-uniform degree social networks, are analyzed with the derivation of closed-form expressions of the stable network diffusion states. Moreover, we analyzed the Erd\H{o}s-R\'enyi random network and the Barab\'asi-Albert scale-free network. To validate our theoretical analysis, we conducted experiments on synthetic networks, real-world Facebook networks, as well we real-world information diffusion dataset of Twitter and Memetracker. All the experiment results were consistent with corresponding theoretical results, which corroborated that our proposed graphical EGT framework is effective and practical for modeling the information diffusion problem.

\bibliographystyle{IEEEtran}

\bibliography{list}
\end{document}